# A Hypothesis for the Speed of Propagation of Light in electric and magnetic fields and the Planning of an Experiment for its Verification.

Wolfenbüttel, November - 23 - 2007

by Claus W. Turtur

## Abstract:

As generally known, the speed of propagation of light in solid state bodies can be different from the speed of light in vacuum. That the mere presence of electric or magnetic fields in the vacuum can suffice to influence the speed of light, is a hypothesis under discussion, which is based on considerations of Quantumelectrodynamics.

For a verification of this hypothesis, an interference-experiment might be performed, of which the planning is given in this article.

## Structure of the article:
1. Theoretical Basis
2. Experimental investigations up to now
3. The hypothesis of the speed of light in electric and magnetic fields
4. Suggestion of a possible experiment

## Contents:

### 1. Theoretical Basis

Already since 1935, nonlinear optical phenomena have been ascribed to the vacuum [Eul 35]. Based on this work, [Rik 00] theoretically analyses the birefringence as an optical property of the vacuum in connection with electric field (Kerr-effect), with magnetic fields (Cotton-Mouton-effect) as well as the magnetoelectric Jones-birefringence. His calculations come to values of $\Delta n = n_\| - n_\perp$ (the directions "$\|$" and "$\perp$" are defined by the applied fields) of $\Delta n_{Kerr} \approx -4.2 \cdot 10^{-41} \cdot E^2$, $\Delta n_{Cotton-Mouton} \approx 3.9 \cdot 10^{-24} \cdot B^2$ and $\Delta n_{Jones} \approx 2.67 \cdot 10^{-32} \cdot E \cdot B$, with $B$ as magnetic field in Tesla and $E$ as electric field in $V/m$. His results are confirmed by others, such as [Bia 70]. The theoretical background is the consideration quantumelectrodynamical (QED) corrections to the Euler-Heisenberg lagrangian in constant electric and magnetic fields because of vacuumpolarisation, i.e. the influence of the fields to the movement of the electrons and positrons within the vacuum. Some important examples are shown in fig.1.

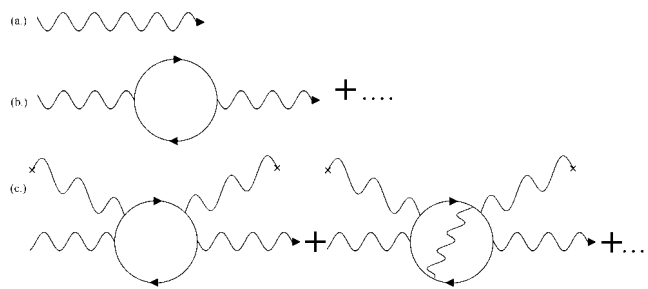

**Fig.1:**
(a.) Classical propagation of the photon in the vacuum without electric or magnetic fields.
(b.) Example for QED corrections without electric or magnetic fields. (virtual $e^+e^-$ – pair).
(c.) Example for additional QED corrections with constant electric or magnetic fields (virtual $e^+e^-$ – pair and radiative corrections).

Furthermore [Boe 02] give a prediction of a reduction of the speed of light in magnetic fields of the amount of $1 - \frac{v}{c} = a \cdot 6.627 \cdot 10^{-25} \cdot B^2 \cdot \sin(\theta)$, where $\theta$ is the angle between the light and the magnetic field, and the constant of $a$ describing the polarization mode of the light, with $a=8$ for the $\|$-mode and $a=14$ for the $\perp$-mode. As we will discus in section 3, there is no disagreement between this value and $\Delta n_{Cotton-Mouton}$ as mentioned above. Regarding the theoretical predictions of the reduction of the speed of light in electric fields we did not find any information in literature.

## 2. Experimental investigations up to now

The experimental verification of these theoretical predictions can not be regarded as already done. [Rik 03] calls all electro- and magneto- optical effects of the vacuum as „far from being experimentally demonstrated". The situation is not changed up to now. At [Riz 07] we find a presentation at a recent meeting where he gives an overview of the experimental situation now. According to this, some elder attempts for the measurement of the rotation of polarisation have been tried without success. Furthermore there is one single experiment [Zav 06] which claimed a result, namely a magnetooptical rotation of polarisation of $(3.9 \pm 0.5) \cdot 10^{-12} \, rad/meter$ (given as angle of the rotation per passed distance in the magnetic field) at a magnetic field of $5.5 \, T$ and a wavelength of light of 1064 nm. In the experimental setup described there, this would mean a birefringence with $\Delta n < 0$. The problem is, that the result of this experiment is in clear contradiction with the theoretical predictions mentioned above, and furthermore it shows some artefacts [Lam 07] which make the probability of a real result rather uncertain. Besides, there is an additional test-experiment of this group [Zav 07], where they withdraw their [Zav 06]-result.

So far, we see no reliable result about the magnetooptical rotation of the polarisation in vacuum. Regarding electrooptical rotation of the polarisation in vacuum, we could not find any experiment up to now, but only some information that an experiment is in the state of being planned, with the aim to investigate light-light-interaction [Sch 07]. Regarding the reduction of the speed of light by electric or magnetic fields in the vacuum, we could not find any experimental information in literature up to now.

## 3. The hypothesis of the speed of light in electric and magnetic fields

Purpose of this article is the suggestion of a novel experiment, which should give the hope to obtain a significant result. Up to now, the experiments tried to find birefringence. Probably it should be more hopeful to analyze the reduction of the speed of light in electric or magnetic fields, because several detectors with extreme good precision to analyze the speed of light already exist: The detectors for gravitational waves. This means, that the aim would not be to find birefringence but to find refraction.

It is clear that every birefringent medium also causes refraction, because in the case of birefringence it is $n_\| - n_\perp = 0 \Rightarrow n_\| \neq n_\perp$, which has the consequence, that $n_\| = n_\perp = 1$ is impossible. So, at least one of both indices $n_\|, n_\perp$ has to be different from 1, which is already enough as a criterion for diffraction (for unpolarized light) – as far as it concerns our experiment being in the phase of planning. For this consideration we did not take any properties of the medium into account, so it should be valid also for vacuum. This is plausible, because in the vacuum as well as in solid state bodies some oscillations of electrically charged particles are responsible for birefringence as well as for refraction.

Typically (in many cases) the difference $\Delta n = n_\| - n_\perp$ is smaller than the absolute distance between $n$ and 1 (see for instance [Hec 05]). This means that $|\Delta n| = |n_\| - n_\perp| < n - 1$, which seems plausible because of the fact, that diffraction is a consequence of the oscillation of charges, and birefringence is based only on the spatial anisotropic part of these oscillations. In our case we can compare the magnetic reduction of the speed of light of $1 - \frac{v}{c} = a \cdot 6.627 \cdot 10^{-25} \cdot B^2 \cdot \sin(\theta)$ with $\Delta n_{Cotton-Mouton} \approx 3.9 \cdot 10^{-24} \cdot B^2$ and find that for $\theta = 90°$, that $a = 8$ leads to $\frac{c}{v} - 1 = 5.3 \cdot 10^{-24}$, and $a = 14$ leads to $\frac{c}{v} - 1 = 9.28 \cdot 10^{-24}$, which means, we find $\frac{c}{v} - 1 = 7.29 \cdot 10^{-24}$ for unpolarized light, which is indeed a bit more than $\Delta n_{Cotton-Mouton}$. The difference is not so large, that it will give rise to a diffraction-experiment much easier than birefringence measurements, but it demonstrates, that an interferometer (which can not measure polarization) does not obstruct the measurement.

So the consequence is: If we find an interferometer with better resolution in $(n-1)$ compared to the $\Delta n$-resolution of the polarimeters used up to now, this might be an advance for the measurement.

As mentioned above, interferometers with very good resolution are detectors for gravitational waves. But they have rather long arms, which are necessary for such high resolution (with a length of several 100 Meters or even several 1000 Meters). Consequently it would be rather hopeless, to apply a strong magnetic field as for instance 10 *Tesla* to such a huge volume. It is much easier to apply an electrostatic field (even at the field strength of vacuum breakdown), because it "only" requires some capacitor plates to be charged. In such a case we could use the following estimation:

An electric field strength at the breakdown of the vacuum of about $E = 500 \frac{V}{\mu m}$ leads to

$$\left(n_{E\,field} - 1\right) > \left| -4.2 \cdot 10^{-41} \cdot E^2 \right| \approx 1.05 \cdot 10^{-23}$$

The mathematic sign ">" shall symbolize the expectation, that $(n_{Feld} - 1)$ should be a bit less tiny than the estimation from the knowledge of $|\Delta n|$, without yet knowing, whether ">" stands for a factor 1,2,3 or 4 (but this is not the main critical point for the planning of an experiment).

## 4. Suggestion of a possible experiment

We want to think about new experimental possibilities, regarding the use of an interferometer in order to verify the ability of an electric field to slow down the speed of light in the vacuum:

● Interferometers with very high resolution, which already provide vacuum, can be found at detectors for gravitational waves. For instance the GEO600-detector has a resolution of $\Delta L/L = 10^{-20}$ on the basis of a Michelson-interferometer [Lig 03]. If it would be possible to introduce an optical resonator, similar as it was done in the [Zav 06]-experiment, who was able to realize about $k = 42000$ passes of the optical path (or even a bit more), the limit of resolution of the GEO600-interferometer would be at $(n-1) = \frac{\Delta L/L}{k} = \frac{10^{-20}}{42000} \stackrel{TR}{\approx} 2.4 \cdot 10^{-25}$. This would be much better than necessary to verify the influence of an electric field on the speed of light.

● Perhaps it is sufficient, to use an instrument which is less difficult available, than a gravitational wave device. In such a case it could be helpful to enhance the number of passes in the optical resonator, because the influence of the field is proportional to the length of lightpath within the field.

<u>Aim of the experiment</u>
1. The principle verification of the hypothesis, that the speed of propagation of light can be influence by electric and magnetic fields, i.e. that the vacuum has not only birefringent properties but also refraction, not only in magnetic fields but also in electric fields.
2. A quantitative statement about the refractive properties of the vacuum.
3. In the language of elementary particle physics a positive result would be regarded as a step towards the optical detection of the axion.

**Author's Adress:**
Prof. Dr. Claus W. Turtur
University of Applied Sciences Braunschweig-Wolfenbüttel
Salzdahlumer Straße 46 / 48
Germany - 38302 Wolfenbüttel
Email: c-w.turtur@fh-wolfenbuettel.de
Tel.: (++49) 5331 / 939 – 3412